\RequirePackage{fix-cm}
\documentclass[smallextended]{svjour3}       
\smartqed  
\usepackage{graphicx}
\usepackage{amsmath}
\usepackage[numbers,sort&compress]{natbib}
\usepackage{bm}
\usepackage{tabularx}
\usepackage{mathptmx}      
\usepackage{slashed}
\usepackage{xcolor}
\usepackage[verbose,hypertexnames=false]{hyperref}
\definecolor{blue}{rgb}{0.0, 0.0, 1.0}
\definecolor{red}{rgb}{1.0, 0.0, 0.0}
\definecolor{royalblue}{rgb}{0.0, 0.14, 0.4}
\hypersetup{colorlinks=true,linkcolor=red, citecolor=blue, urlcolor=blue}
\def\orcid#1{\kern .08em\href{https://orcid.org/#1}{\includegraphics[keepaspectratio,width=0.7em]{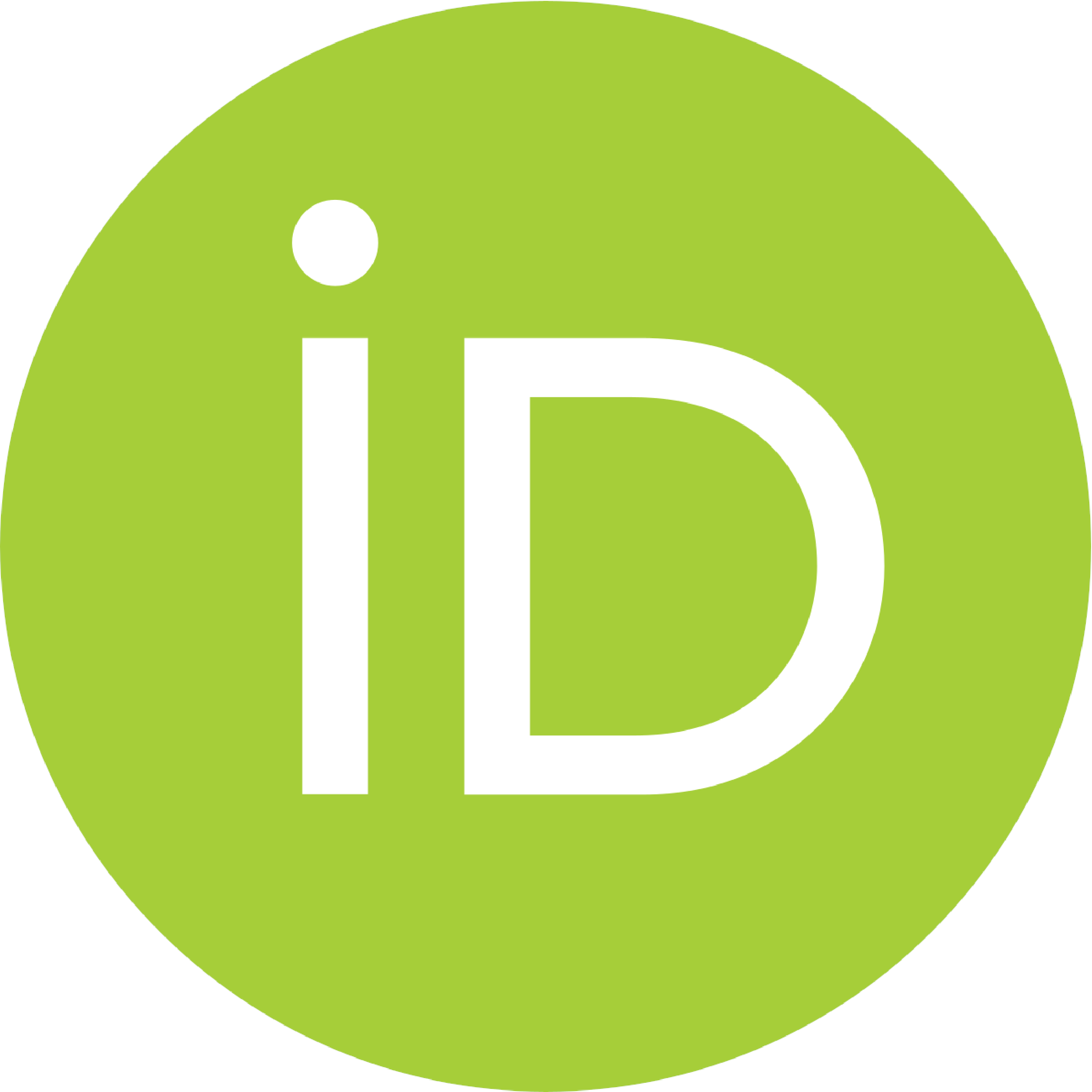}}}%
\journalname{Few-Body Systems}
\begin{document}

\title{New Elementary Operator for Kaon Photoproduction on the Nucleon and Nuclei
}


\author{Terry Mart\orcid{0000-0003-4628-2245}         \and
        Jovan Alfian Djaja 
}


\institute{Terry Mart$^\dagger$ and Jovan Alfian Djaja$^\S$ \at
              Departemen Fisika, FMIPA, Universitas Indonesia, Depok 16424, Indonesia \\
              \email{$^\dagger$terry.mart@sci.ui.ac.id, $^\S$jovan.alfian@ui.ac.id}           
}

\date{Received: date / Accepted: date}

\maketitle

\begin{abstract}
A new elementary operator for kaon photoproduction on the nucleon and nuclei has been developed within a Feynman diagrammatic framework. By fitting the unknown coupling strengths at the electromagnetic and hadronic vertices of the baryon resonances to all available experimental data across the six isospin channels, the model achieves excellent agreement with the data. The operator includes 26 nucleon resonances in the $K\Lambda$ channels and 17 additional $\Delta$ resonances in the $K\Sigma$ channels. For applications to nuclear reactions, such as hypernuclear photoproduction, the operator is formulated in Pauli space, allowing a straightforward implementation of the nonrelativistic approximation. Several alternative forms for expressing the operator output are proposed. In one of them, the spin operators and photon polarization vectors are separated from the operator, since both are frame dependent, thereby enhancing its versatility in nuclear applications.
\end{abstract}

\section{Introduction}
\label{intro}
Two of the most fundamental questions in particle physics concern the nature of the building blocks of the universe and the mechanisms governing their interactions. Remarkably, even after more than a century of research, these questions remain only partially answered. The current theoretical framework, the Standard Model, has achieved great success but still faces several unresolved issues. One of the major challenges lies in understanding the strong interaction in its non-perturbative regime, which continues to be a central focus of particle and nuclear physics. This area of study has evolved into a distinct and vibrant field known as hadronic physics.

At the core of hadronic physics lies the nucleon–nucleon ($NN$) interaction. To extend this framework toward a more unified description of baryon–baryon interactions, as suggested by SU(3) flavor symmetry, it is essential to incorporate the hyperon-nucleon ($YN$) interaction via the strangeness degree of freedom. A key process that explicitly involves strangeness and provides valuable insight into this extension is kaon photoproduction, in which a hyperon ($Y$) is simultaneously produced to conserve strangeness. A theoretical understanding of this process requires detailed knowledge of both the electromagnetic interactions, $\gamma NN$, $\gamma YY$, and $\gamma KK$, and the strong interaction $KYN$. With support from experimental data, these interactions can be studied within phenomenological frameworks such as isobar or constituent quark models.

Kaon photoproduction on the nucleon has been investigated since the late 1950s \cite{Moravcsik:1957}. Surprisingly, this early work already considered all six possible isospin channels listed in Table~\ref{tab:threshold_energy}. However, the lack of experimental data and detailed knowledge of the relevant particles at that time inevitably led to limited results. For instance, the cross sections for the $\gamma n \to K^0 \Lambda$ and $\gamma n \to K^0 \Sigma^0$ channels were predicted to be zero, since only the Born terms were included and neither the nucleon anomalous magnetic moments nor resonance contributions were taken into account due to the absence of information on these parameters.

Nearly a decade later, a more comprehensive phenomenological study of the $\gamma p \to K^+ \Lambda$ channel became possible, supported by sufficient experimental data to draw more definitive conclusions \cite{Thom:1966rm}. Since then, numerous theoretical \cite{Adelseck:1985scp,Adelseck:1990ch,Mart:1995wu,David:1995pi,Mart:1999ed,Maxwell:2004ga,Mart:2019mtq} and experimental \cite{SAPHIR:1998fev,SAPHIR:1999wfu,CLAS:2005lui,CLAS:2006pde,CBELSATAPS:2007oqn,GRAAL:2008jrm,CLAS:2009rdi,CLAS:2009fmu,CLAS:2010aen,CrystalBallatMAMI:2013iig,CLAS:2016wrl,CLAS:2017gsu} studies have significantly deepened our understanding of this process. Although the primary motivation for these works has been to elucidate the underlying mechanisms from a particle-physics perspective, kaon photoproduction also plays an important role in nuclear physics, particularly in the study of hypernuclear production. This aspect forms the main focus of the present work, namely, the construction of an elementary operator that encapsulates the dynamics of the elementary reaction at the particle level for use in nuclear calculations.

\begin{table}[t]
\caption{Threshold energies for kaon photoproduction on the nucleon, $\gamma N \to K Y$, for the six possible isospin channels, expressed in terms of the photon laboratory energy $E_\gamma^{\rm thr}$ and the total c.m. energy $W^{\rm thr}$.}
\label{tab:threshold_energy}       
\begin{tabularx}{\textwidth} { 
   >{\raggedright\arraybackslash}X 
   >{\centering\arraybackslash}X 
   >{\centering\arraybackslash}X  }
\hline\noalign{\smallskip}
Channel & $E_\gamma^{\rm thr}$ [MeV] & $W^{\rm thr}$ [MeV] \\[0.1ex]
\noalign{\smallskip}\hline\noalign{\smallskip}
$\gamma p \to K^+ \Lambda$ & ~\,911.1 & 1609.4  \\
$\gamma n \to K^0 \Lambda$ & ~\,915.3 & 1613.3  \\
$\gamma p \to K^+ \Sigma^0$ & 1046.2 & 1686.3 \\
$\gamma p \to K^0 \Sigma^+$ & 1047.4 & 1687.0 \\
$\gamma n \to K^0 \Sigma^0$ & 1050.6 & 1690.3  \\
$\gamma n \to K^+ \Sigma^-$ & 1052.1 & 1691.1 \\
\noalign{\smallskip}\hline
\end{tabularx}
\end{table}

The elementary operator must account for all possible isospin channels, as possible rescattering processes require the inclusion of channels beyond the one directly considered. Table~\ref{tab:threshold_energy} lists the six possible isospin channels of kaon photoproduction along with their corresponding threshold energies, expressed in terms of the photon laboratory and total c.m. energies. Unlike pion photoproduction, the relatively high thresholds in kaon photoproduction allow contributions from numerous baryon and meson resonances even near the production threshold. This circumstance makes the formulation of kaon photoproduction inherently more complex than that of pion photoproduction. As a result, the photoproduction of kaons continues to attract significant interest as an active subject of study.

This paper is organized as follows. Section~\ref{sec:on_nucleon} presents the basic formalism of the elementary operator for kaon photoproduction on the nucleon and compares the calculated observables with the available experimental data. Section~\ref{sec:nuclear_application} describes the construction of the operator and its application to nuclear processes. Finally, Sec.~\ref{sec:conclusion} summarizes this work and provides the conclusion.

\section{Kaon Photoproduction on the Nucleon}
\label{sec:on_nucleon}
\subsection{Formalism}
\begin{figure}[htbp]
  \includegraphics[width=1.0\columnwidth]{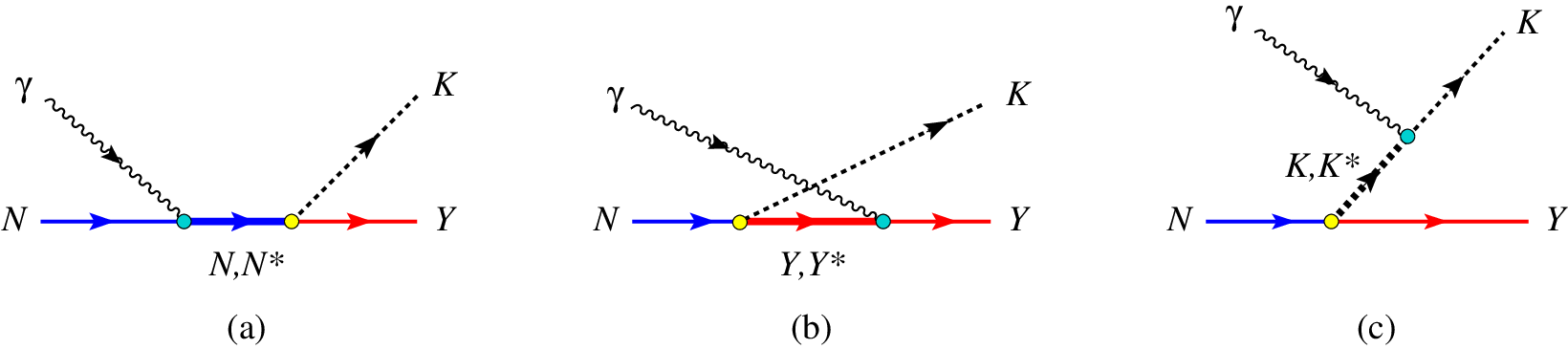}
\caption{Feynman diagrams for kaon photoproduction on the nucleon, showing the contributions from (a) $s$-channel intermediate nucleon and nucleon-resonance states, (b) $u$-channel intermediate hyperon and hyperon-resonance states, and (c) $t$-channel intermediate kaon and kaon-resonance states.}
\label{fig:feynman}       
\end{figure}

In general, kaon photoproduction on the nucleon in six isospin channel can be written as
\begin{equation}
    \label{eq:reaction}
    \gamma(k)+N(p_N)\to K(q)+Y(p_Y) ,
\end{equation}
The corresponding Feynman diagrams are shown in Fig.~\ref{fig:feynman}, illustrating the contributions from the Born terms, which involve intermediate nucleon, hyperon, and kaon states, as well as from their respective resonance states. It follows from Eq.~(\ref{eq:reaction}) that the Mandelstam variables can be expressed as
\begin{eqnarray}
    \label{eq:mandelstam}
	s=(k+p_N)^2, ~~~ u=(k-p_Y)^2, ~~~ {\rm and}~~~  t=(k-q)^2 .
\end{eqnarray}
Concerning the future extension of the operator to include electroproduction, from now on we shall consider the formalism of kaon electroproduction, which is more general, and indicate the necessary modifications when reducing it to the photoproduction case.

The vertex factors and propagators for the Born terms are well established and can be found, for example, in Ref.~\cite{Adelseck:1985scp,Adelseck:1990ch,Mart:2010ch}. Since hadronic form factors are included at the hadronic vertices, gauge invariance of the Born terms must be restored. For this purpose, we adopt the prescription of Haberzettl~\cite{Haberzettl:1998aqi}. The resonance terms are, by construction, gauge invariant.

For nucleon resonances, the corresponding propagators and vertex factors are well established only for spin-$1/2$ states. For higher-spin resonances, ambiguities arise, particularly due to the presence of lower-spin background contributions. To eliminate these unphysical components in the scattering amplitude, we employ the consistent interaction formalism proposed by Pascalutsa~\cite{Pascalutsa:2000kd} and explicitly derived later by Vrancx et al.~\cite{Vrancx:2011qv}. Within this framework, the scattering amplitude for nucleon resonances with mass $m_R$ and spin $n+\frac{1}{2}$ can be written as~\cite{Luthfiyah:2021yqe} 
\begin{eqnarray}\label{scatt}
\mathcal{M}_{\mathrm{res}}^{n+1/2} &=&\bar{u}_Y\, \prod_{i=1}^{n}\slashed{p}_{R}\,g_{\alpha_i\,\mu_i}\tilde{\Gamma}_{\rm had}^{\alpha_1...\alpha_n}\,\frac{1}{m_{R}^{2n}}\,\frac{\slashed{p}_{R} + m_{R}}{p_{R}^2 - m_{R}^2 + im_{R}\Gamma } 
 \mathcal{P}^{\mu_1...\mu_n,\nu_1\ldots\nu_n}_{(n+1/2)}\,(p_{R})\nonumber\\ &&\times\prod_{i=1}^{n}\slashed{p}_{R}\,g_{\beta_j\,\nu_j}\tilde{\Gamma}_{\rm em}^{\beta_1\ldots\beta_n}\,u_p \nonumber \\
&=&\bar{u}_Y\,\tilde{\Gamma}^{\rm had}_{\mu_1\dots\mu_n}\,\frac{p_R^{2n}}{m_{R}^{2n}}\,\frac{\slashed{p}_{R} + m_{R}}{p_{R}^2 - m_{R}^2 + im_{R}\Gamma}\,\mathcal{P}^{\mu_1\dots\mu_n,\nu_1\ldots\nu_n}_{(n+1/2)}\,(p_{R})\,\tilde{\Gamma}_{\rm em}^{\beta_1\ldots\beta_n}\,u_p \, ,
\end{eqnarray}
where $\tilde{\Gamma}^{\rm had}_{\mu_1\ldots\mu_n}$ and $\tilde{\Gamma}^{\rm em}_{\beta_1\ldots\beta_n}$ denote the hadronic and electromagnetic vertex factors, respectively, while $\mathcal{P}^{\mu_1\ldots\mu_n,\nu_1\ldots\nu_n}_{(n+1/2)}(p_{R})$ represents the on-shell projection operator for a spin-$\left(n+\frac{1}{2}\right)$ nucleon resonance.

As an example, the hadronic and electromagnetic vertices for spin-${15}/{2}$ nucleon resonance are given by~\cite{Luthfiyah:2021yqe} 
\begin{eqnarray}
  \label{eq:had_15}
\lefteqn{\Gamma_{\rm{had}}^{\mu\mu_1\mu_2\mu_3\mu_4\mu_5\mu_6\pm} ~=~ }\nonumber\\[0.7ex] && \frac{g_{KYR}}{m_{R}^8}\Gamma_{\pm}\bigl[(p_Y\cdot q - \slashed{p}_Y\slashed{q})\gamma^\mu + \slashed{p}_Y q^{\mu} - \slashed{q}p^{\mu}_Y)\bigr]
p^{\mu_1}_Y p^{\mu_2}_Y p^{\mu_3}_Y p^{\mu_4}_Y p^{\mu_5}_Y p^{\mu_6}_Y ~,
\end{eqnarray}
and
\begin{eqnarray}
  \label{eq:em_15}
\Gamma_{\rm{em}}^{\nu\nu_1\nu_2\nu_3\nu_4\nu_5\nu_6\pm} &=& \frac{-i}{m_{R}^8} \biggl[g_1p^{\nu}(\slashed{k}\slashed{\epsilon} -  \slashed{\epsilon}\slashed{k})+ g_2(k^{\nu}p\cdot \epsilon - \epsilon^{\nu}p\cdot k)
+g_3 (\epsilon^\nu\slashed{k} - k^\nu\slashed{\epsilon})\slashed{p}   
\nonumber\\
&& +g_4 \gamma^{\nu} (\slashed{k}\slashed{\epsilon} -  \slashed{\epsilon}\slashed{k})\slashed{p} + g_5\gamma^{\nu}(p\cdot\epsilon\slashed{k} - p\cdot k\slashed{\epsilon})\biggr]
k^{\nu_1}k^{\nu_2}k^{\nu_3}k^{\nu_4}k^{\nu_5}k^{\nu_6}\Gamma_{\mp}\, ,\nonumber\\
\end{eqnarray}
respectively, where the parity factor $\Gamma_{+} = i\gamma_5$ and $\Gamma_{-} = 1$, for nucleon resonances with positive and negative parities, respectively. The products of coupling constants $g_{KYR}~ g_i\equiv G_{KYR}^{(i)}$ are determined by fitting the calculated observables to the available experimental data.

In the case of electroproduction, the total scattering amplitude, which is obtained by summing the contributions from all Born and resonance diagrams, can be decomposed into six gauge- and Lorentz-invariant matrices,
\begin{eqnarray}
\label{eq:M_fi_defined_Ai_Mi}
\mathcal{M}_{\mathrm{fi}}=\bar{u}_Y\sum^{6}_{i=1} A_i(s,t,u,k^2)\, M_i\,\, u_N ~.
\end{eqnarray}
The six invariant matrices $M_i$ are given by \cite{Mart:2015jof,Mart:2008gq}
\begin{subequations}
\label{eq:Ai_Mi}
\begin{align}
M_1 ~&=~ \tfrac{1}{2}\gamma_5\left(\slashed{\epsilon}\slashed{k}-\slashed{k}\slashed{\epsilon}\right),\\
M_2 ~&=~ \gamma_5\left[\left(2q-k\right)\cdot \epsilon\, P\cdot k-\left(2q-k\right)\cdot k\, P\cdot \epsilon\right], \\
M_3 ~&=~ \gamma_5\left(q\cdot k\, \slashed{\epsilon}-q\cdot\epsilon\,\slashed{k}\right),\\
M_4 ~&=~ i\varepsilon_{\mu\nu\rho\sigma}\gamma^\mu q^\nu\epsilon^\rho k^\sigma,\\
M_5 ~&=~  \gamma_5 \left(q \cdot \epsilon \, k^2 - q \cdot k \, k \cdot \epsilon \right),\\
M_6 ~&=~  \gamma_5 \left(k \cdot \epsilon \, \slashed{k} - k^2 \epsilon \right),
\end{align}
\end{subequations}
where $\epsilon_\mu$ denotes the photon polarization vector, $P \equiv (p_N + p_Y)/2$, and $\varepsilon_{\mu\nu\rho\sigma}$ is the four-dimensional Levi-Civita tensor. In the case of photoproduction, we have $k^2 = k \cdot \epsilon = 0$, and consequently, the longitudinal amplitudes $M_5$ and $M_6$ vanish.

The invariant amplitudes $A_i$ extracted from Eq.~(\ref{eq:M_fi_defined_Ai_Mi}) can be employed to calculate all experimentally measured observables. A comprehensive list of complete observables for meson photo- and electroproduction is provided, for example, in Ref.~\cite{Knochlein:1995qz}, where the cross sections and single- or double-polarization observables are expressed in terms of the Dennery~\cite{Dennery:1961zz} or CGLN~\cite{Chew:1957tf} amplitudes $F_1,\ldots,F_6$ in Pauli space. These amplitudes are related to the invariant amplitude of Eq.~(\ref{eq:M_fi_defined_Ai_Mi}) through
\begin{eqnarray}
\label{eq:Mi_to_Fi}
    \mathcal{M}_{fi}=
    {\bar u}(p_Y)\,\sum_{i=1}^{6}A_i(s,t,u,k^2)\, M_i\, u(p_N) &=& \chi^\dagger_{Y}~{\cal F}~\chi_{N} \, ,
\end{eqnarray}
where 
\begin{eqnarray}
\label{eq:F_in_terms_of_F1_F6}
    {\cal F} &=& i\bm{\sigma\cdot a}\, F_1+\bm{\sigma\cdot}\bm{\hat q}\,\bm{\sigma\cdot}(\bm{\hat k}\bm{\times a})\,F_2+i\bm{\sigma\cdot}\bm{\hat k}\,\bm{\hat q}\bm{\cdot a}\,F_3 \nonumber\\&& +~ i\bm{\sigma\cdot}\bm{\hat q}\,\bm{\hat q}\bm{\cdot a}\,F_4 +i\bm{\sigma\cdot}\bm{\hat k}\,\bm{\hat k}\bm{\cdot a}\,F_5 + +i\bm{\sigma\cdot}\bm{\hat q}\,\bm{\hat k}\bm{\cdot a}\,F_6 \, ,
\end{eqnarray}
with 
\begin{equation}
    a_\mu = \epsilon_\mu - \epsilon_0 k_\mu /k_0 \,.
\end{equation}
By substituting Eqs.~(\ref{eq:Ai_Mi}) and (\ref{eq:F_in_terms_of_F1_F6}) into Eq.~(\ref{eq:Mi_to_Fi}) we obtain
\begin{subequations}
\begin{align}
    F_{1,2} ~&=~ \frac{1}{8\pi W} \sqrt{\left(E_N\pm m_N\right)\left(E_Y\pm m_Y\right)} \Bigl\{ \pm(W\mp m_N) A_1+ k\cdot q\, (A_3-A_4) \nonumber\\
    ~&~ +(W\mp m_N)(W\mp m_Y) A_4 -k^2 A_6 \Bigr\}
    ,\\
    F_{3,4} ~&=~ \frac{|\bm{k}| |\bm{q}|}{8\pi W} 
    \sqrt{\frac{E_Y\pm m_Y}{E_N\pm m_N}} \Bigl\{ (s-m_N)^2 A_2 \mp
    {\textstyle \frac{1}{2}} k^2(A_2-A_5) + (W\pm m_N) (A_3-A_4) \Bigr\} .
    \\
    F_{5,6} ~&=~ \frac{k_0}{8\pi W} \sqrt{\left(E_N\pm m_N\right)\left(E_Y\pm m_Y\right)} \Bigl[ \pm A_1 + (W\mp m_Y) A_4 - (W\mp m_N) A_6
    \nonumber\\ ~&~ -\frac{1}{E_N\pm m_N}\, \Bigl\{ \pm W(|\bm{k}|^2-2 \bm{k\cdot q}) A_2 \pm (k\cdot q\, k_0-E_K k^2) (A_5-{\textstyle \frac{3}{2}} A_2)\nonumber\\
    ~&~ - [E_K(W\pm m_N)-k\cdot q](A_3-A_4) \Bigr\} \Bigr] .
\end{align}
\end{subequations}
The above relations can be used to calculate all observables in kaon photo- and electroproduction on the nucleon listed in Ref.~\cite{Knochlein:1995qz}. At this stage, it is important to note that the Pauli amplitudes $F_5$ and $F_6$ vanish in the case of photoproduction.

\subsection{Comparison with Experimental Data}
A reliable elementary operator for use in nuclear reactions must be capable of accurately describing the underlying elementary process. Therefore, before discussing the formalism within the nuclear environment, it is essential to examine the performance of the operator in reproducing experimental data at the elementary (or particle) level. The performance of the isobar model employed to construct the present elementary operator has been comprehensively analyzed in Ref.~\cite{Luthfiyah:2021yqe}. Here, we provide a brief overview of the comparison between the calculated observables and the available experimental data. In addition, we compare the results of the present operator with those predicted by Kaon-Maid to evaluate the degree of improvement achieved with the current elementary operator.

It is important to note that for both the $K\Lambda$ and $K\Sigma$ channels, two models were discussed in Ref.~\cite{Luthfiyah:2021yqe}. For the $K\Lambda$ channels, Model A was constructed by including 23 nucleon resonances with spins up to $13/2$ and fitting the calculated observables to 9364 experimental data points, resulting in a $\chi^2$ per degree of freedom ($\chi^2/N_{\rm dof}$) of 1.46. Model B, taken from an earlier study \cite{Mart:2019mtq}, used 21 nucleon resonances and the same experimental database, yielding a $\chi^2/N_{\rm dof}=1.52$. For the $K\Sigma$ channels, two corresponding models were also considered. Model C employs the same set of nucleon resonances as in Model A and, in addition, includes 17 $\Delta$ resonances. By fitting to approximately 8000 data points, Model C yields a $\chi^2/N_{\rm dof}=1.18$. Model D uses 21 nucleon and 14 $\Delta$ resonances and was fitted to the same $K\Sigma$ database, resulting in a $\chi^2/N_{\rm dof}=1.22$ \cite{Clymton:2021wof}.

\begin{figure}[htbp]
  \includegraphics[width=1.0\columnwidth]{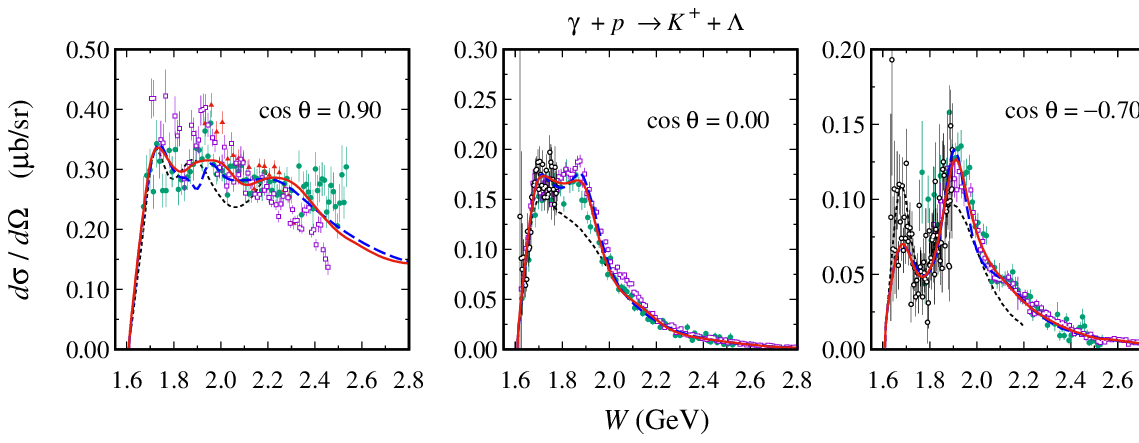}
\caption{Differential cross section for the $\gamma p \to K^+ \Lambda$ reaction as a function of the c.m. energy for three different kaon angles. The solid and dashed curves correspond to Model A and Model B of Ref.~\cite{Luthfiyah:2021yqe}, respectively, while the dotted curves represent the results of Kaon-Maid \cite{Mart:1999ed,Bennhold:1999mt}, which is valid only up to $W=2.2$ GeV. Experimental data are taken from the Crystal Ball 2014 (open circles) \cite{CrystalBallatMAMI:2013iig}, CLAS 2010 (solid circles) \cite{CLAS:2009rdi}, CLAS 2006 (open squares) \cite{CLAS:2005lui}, and LEPS 2006 (solid triangles) \cite{LEPS:2005hji}.}
\label{fig:dif_kpl}       
\end{figure}

Figure~\ref{fig:dif_kpl} shows the comparison between the differential cross sections obtained from Models A and B with those from Kaon-Maid and the experimental data for the $K^+\Lambda$ channel at three kaon angles: forward, right-angle, and backward directions. The amount of available experimental data is evidently abundant. However, at the forward angle, the data exhibit some inconsistencies. Although the CLAS 2006 results appear consistent with those of LEPS 2006, both sets differ from the CLAS 2010 measurement. Consequently, Models A and B reproduce an average trend of these data. This is rather unfortunate, since in nuclear calculations the cross section at forward angles is typically the largest and therefore provides the most favorable kinematics for experiments, given that nuclear cross sections are usually extremely small. Furthermore, at all kaon angles shown in Fig.~\ref{fig:dif_kpl}, two distinct peaks are consistently observed at $W \approx 1.650$ and $1.900~\text{GeV}$. The first peak originates from several nucleon resonances with masses around $1.650~\text{GeV}$, whereas the second arises from the $P_{13}(1900)$ state, as exhibited in Fig.~9 of Ref.~\cite{Mart:2019mtq}. The comparison with Kaon-Maid predictions clearly demonstrates that the present elementary operator yields a substantial improvement in describing the data.

\begin{figure}[htbp]
  \includegraphics[width=1.0\columnwidth]{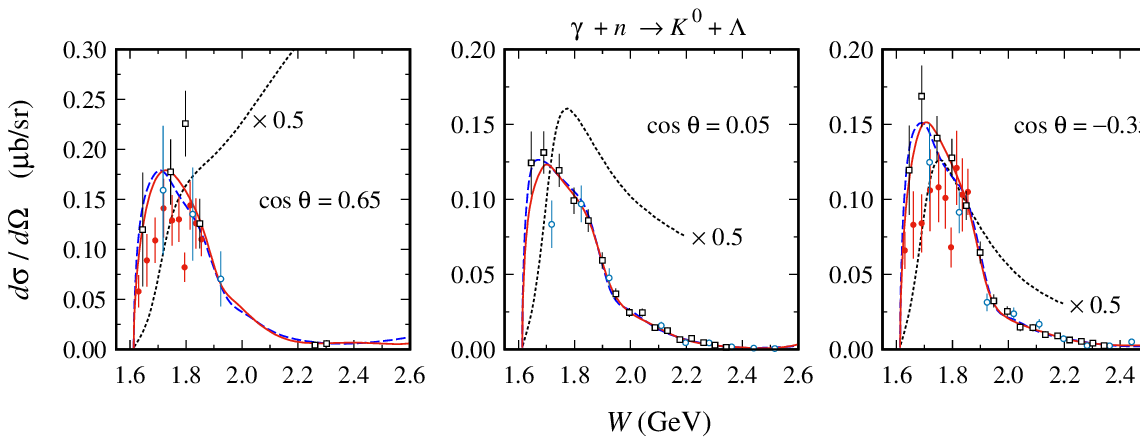}
\caption{Same as Fig.~\ref{fig:dif_kpl}, but for the $\gamma n \to K^0 \Lambda$ reaction. Experimental data are taken from the CLAS g10 and g13 Collaborations (open circles and open squares, respectively) \cite{CLAS:2017gsu}, and from the MAMI 2018 Collaboration (solid circles) \cite{A2:2018doh}. Note that the Kaon-Maid prediction has been multiplied by a factor of 0.5 to fit the scale.}
\label{fig:dif_k0l}       
\end{figure}

Figure~\ref{fig:dif_k0l} highlights the substantial improvement achieved by the present elementary operator in comparison with Kaon-Maid. Nevertheless, this difference is understandable, as the Kaon-Maid results shown in Fig.~\ref{fig:dif_k0l} are purely predictive and obtained solely by invoking isospin symmetry. This contrasts with the results of Models A and B, which were fitted to the currently available experimental data.

\begin{figure}[htbp]
  \includegraphics[width=1.0\columnwidth]{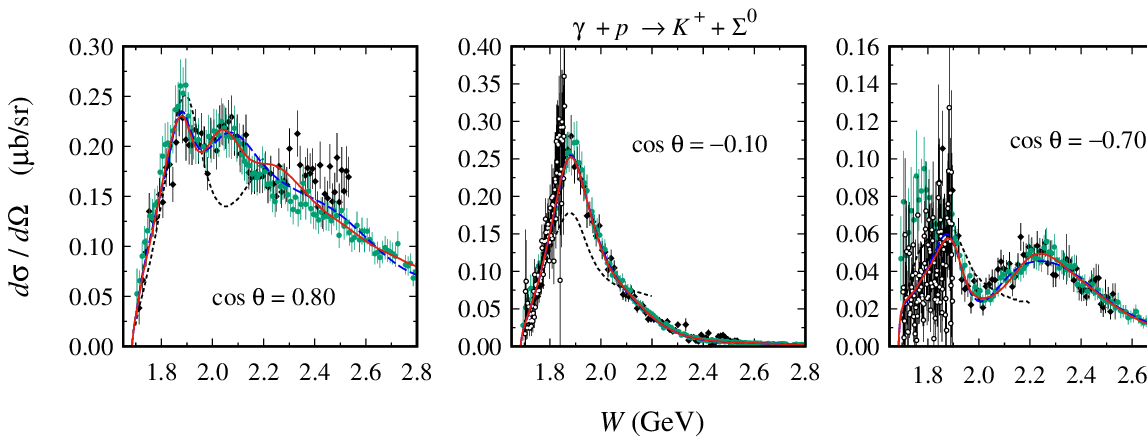}
\caption{Differential cross section for for the $\gamma p \to K^+ \Sigma^0$ channel as a function of the c.m. energy for three different kaon angles. The solid and dashed curves correspond to Model C and Model D of Ref.~\cite{Luthfiyah:2021yqe}, respectively, while the dotted curves represent the results of Kaon-Maid \cite{Mart:2000jv}. Experimental data are taken from the CLAS 2006 (solid diamonds) \cite{CLAS:2005lui}, CLAS 2010 (solid circles) \cite{CLAS:2010aen}, and Crystal Ball 2014 (open circles) \cite{CrystalBallatMAMI:2013iig} Collaborations.}
\label{fig:dif_kps0}       
\end{figure}

\begin{figure}[htbp]
  \includegraphics[width=1.0\columnwidth]{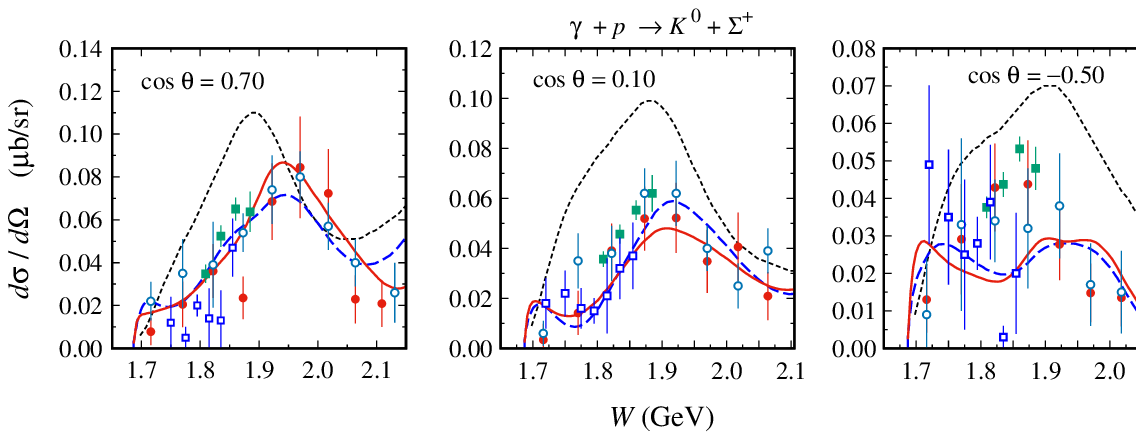}
\caption{As in Fig.~\ref{fig:dif_kps0}, but for the $\gamma+p\to K^0+\Sigma^+$ channel. Experimental data are taken from the SAPHIR 2005 (open circles) \cite{Lawall:2005np}, CBELSA 2008 (solid circles) \cite{CBELSATAPS:2007oqn}, MAMI A2 2013 (solid squares) \cite{A2:2013cqk}, MAMI A2 2019 (open squares) \cite{A2:2018doh}.}
\label{fig:dif_k0sp}       
\end{figure}

\begin{figure}[htbp]
  \includegraphics[width=1.0\columnwidth]{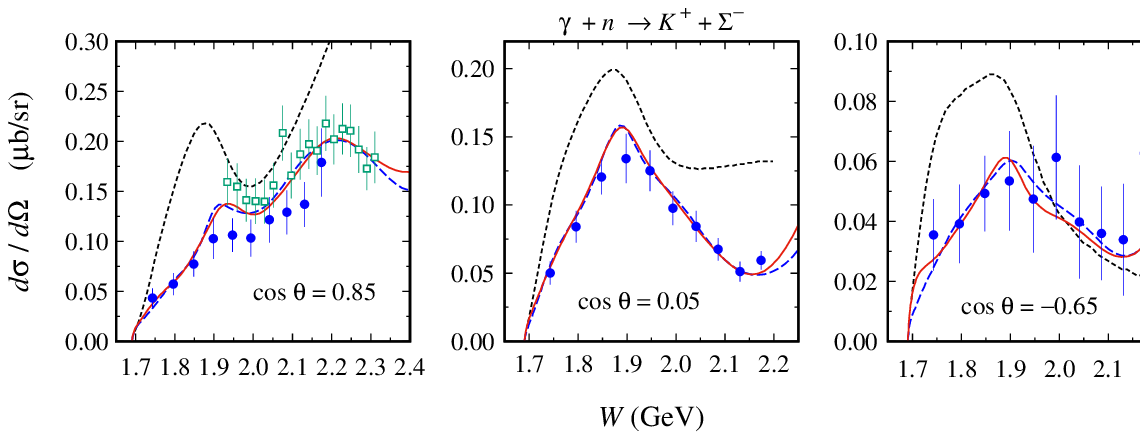}
\caption{As in Fig.~\ref{fig:dif_kps0}, but for the $\gamma+n\to K^++\Sigma^-$ channel. Experimental data are taken from the LEPS 2006 (open squares) \cite{Kohri:2006yx} and CLAS 2010 (solid circles) \cite{CLAS:2009fmu} Collaborations.}
\label{fig:dif_kpsm}       
\end{figure}

\begin{figure}[htbp]
  \includegraphics[width=1.0\columnwidth]{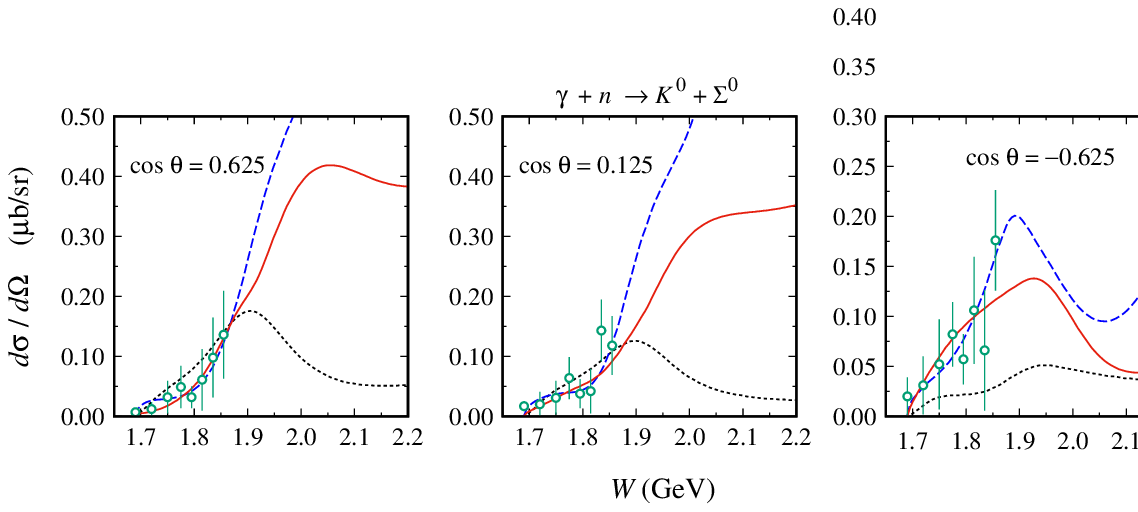}
\caption{As in Fig.~\ref{fig:dif_kps0}, but for the $\gamma+n\to K^0+\Sigma^0$ channel. Experimental data are obtained from the MAMI A2 2019 \cite{A2:2018doh} Collaboration.}
\label{fig:dif_k0s0}       
\end{figure}

In the case of four $K\Sigma$ channels, shown in Figs.~\ref{fig:dif_kps0}-\ref{fig:dif_k0s0}, we can see that experimental data from different collaborations are relatively consistent each other. Again, we see significant improvement has been achieved by the new elementary models. In general, for the kinematics where experimental data are available, the difference between models C and D is insignificant. However, this is not the case when the constraint from experimental data is absent, as shown by the $K^0\Sigma^0$ channel in Fig.~\ref{fig:dif_k0s0}. In this case, all panels, that describe different kaon angles, show that the discrepancy between the two models starts to appear at $W\approx 1.85$ GeV. 

The results presented in Figs.~\ref{fig:dif_kpl}-\ref{fig:dif_k0s0} indicate the energy range within which the present elementary model remains valid. For the $K^+\Lambda$, $K^0\Lambda$, and $K^+\Sigma^0$ channels, the model provides a reliable description from threshold up to $W \approx 2.8$~GeV. In contrast, its validity extends only up to 2.1 and 2.2~GeV for the $K^0\Sigma^+$ and $K^+\Sigma^-$ channels, respectively. Finally, for the $K^0\Sigma^0$ channel, the model remains applicable up to $W \approx 1.9$~GeV, beyond which noticeable deviations begin to appear. 

Since polarization observables are particularly sensitive to the underlying dynamics of a model, it is crucial to examine the performance of the present operator in this regard as well. For brevity, we limit our discussion to the $K\Lambda$ channel at selected total c.m. energies. For a more comprehensive comparison with experimental data, the reader is referred to Refs.~\cite{Mart:2019mtq,Luthfiyah:2021yqe,Clymton:2021wof}. 

\begin{figure}[htbp]
  \includegraphics[width=1.0\columnwidth]{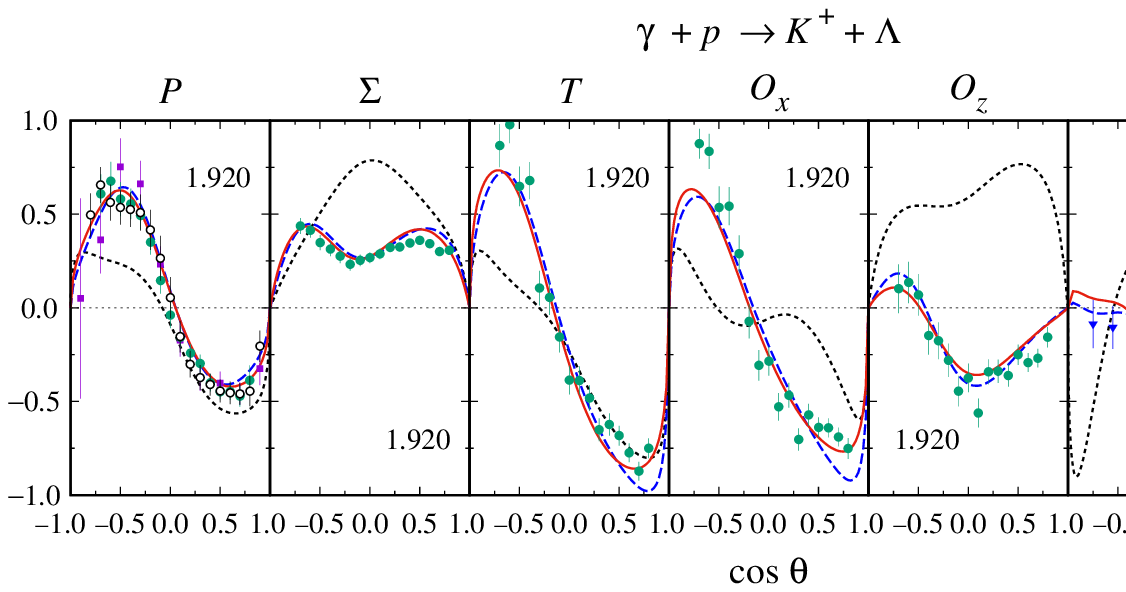}
\caption{Single and double polarization observables for the $\gamma p \to K^+ \Lambda$ reaction as functions of the kaon angle. The corresponding total c.m. energy (in GeV) is indicated in each panel. The notation of the curves is the same as in Fig.~\ref{fig:dif_kpl}. Experimental data are taken from the CLAS 2016 (solid circles) \cite{CLAS:2016wrl}, CLAS 2010 (open circles) \cite{CLAS:2009rdi}, CLAS 2006 (solid squares) \cite{CLAS:2005lui}, and CLAS 2007 (solid triangles) \cite{CLAS:2006pde} Collaborations.}
\label{fig:polarization}       
\end{figure}

Figure~\ref{fig:polarization} presents a comparison of the three phenomenological models discussed above in terms of the single-polarization observables $P$, $\Sigma$, and $T$, as well as the double-polarization observables $O_x$, $O_z$, $C_x$, and $C_z$. It is evident that a similar trend is observed here, with Models~A and~B showing substantial improvements over Kaon-Maid, although small discrepancies with the experimental data are noticeable for the beam–recoil polarization $C_x$.

Based on the performance of the elementary models discussed in this section, Model~A is adopted to describe the $K\Lambda$ channels, while Model~C is used for the $K\Sigma$ channels in constructing the elementary operator for nuclear applications.

\section{Elementary Operator for Nuclear Applications}
\label{sec:nuclear_application}
\subsection{Construction of the Operator}
\begin{figure}[htbp]
  \includegraphics[width=1.0\columnwidth]{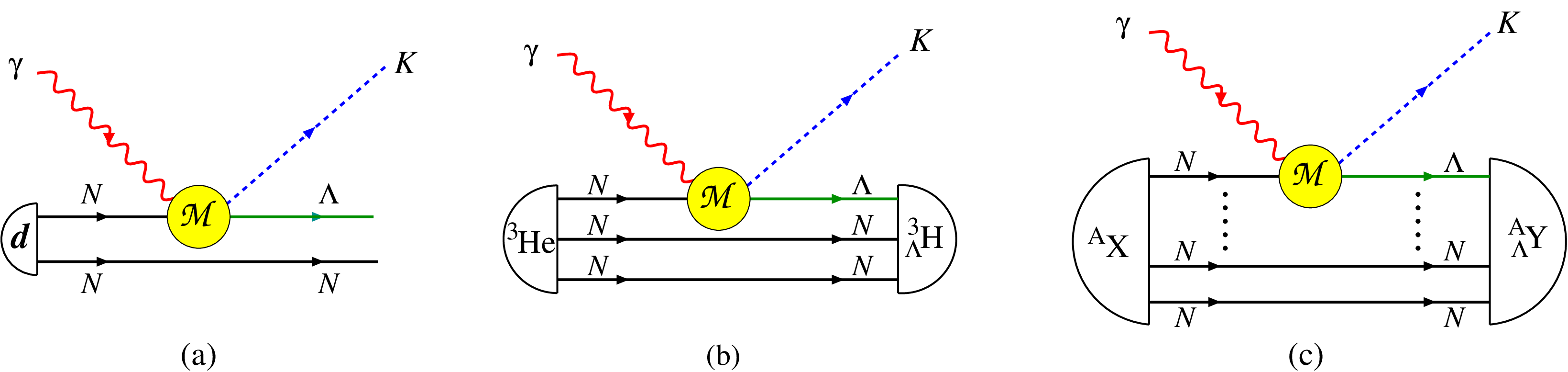}
\caption{Feynman diagrams for kaon photoproduction on (a) the deuteron, leading to an unbound kaon-hyperon final state; (b) helium-3, resulting in the formation of a hypertriton; and (c) a heavier nucleus, producing a heavier hypernucleus. The elementary production operator is denoted by ${M}$ in each diagram.}
\label{fig:feynman_hyp}       
\end{figure}

The photoproduction of kaons on the deuteron, $^3$He, and heavier nuclei can be illustrated by the Feynman diagrams shown in Fig.~\ref{fig:feynman_hyp}, panels (a), (b), and (c), respectively. For the deuteron case, represented in panel (a), we assume the absence of final-state interactions (FSI) for simplicity. As a consequence, the initial nucleon that interacts with the elementary operator is off-shell, whereas the produced hyperon is on-shell. This situation differs from the hypertriton photoproduction depicted in panel (b), where both the initial and final baryons participating in the interaction with the elementary operator are off-shell. A similar condition applies to the production of heavier hypernuclei, as shown in panel (c). Therefore, it is essential to construct a more versatile operator in which the four-momenta of all interacting particles are defined in a frame-independent manner. This is particularly important, since nuclear observables are usually calculated in the nuclear rest frame.

On the other hand, there are several representations of the output of the elementary operator that can be employed to incorporate the dynamics of the elementary reaction into nuclear process calculations. The first representation is in terms of the $A_i$ amplitudes, as defined in Eq.~(\ref{eq:M_fi_defined_Ai_Mi}), which is the most straightforward form. This representation was used, for example, in Ref.~\cite{Salam:2004gz} to investigate kaon–nucleon and hyperon–nucleon interactions in kaon photoproduction on the deuteron. The second option is to use the Dennery or CGLN amplitudes in Pauli space, as given in Eq.~(\ref{eq:F_in_terms_of_F1_F6}). 

The third alternative is to expand the scattering matrix of Eq.~(\ref{eq:M_fi_defined_Ai_Mi}) into the complete set of amplitudes in Pauli space, rather than adopting the Dennery or CGLN forms, i.e., \cite{Mart:2008gq}
\begin{eqnarray}
\lefteqn{ \mathcal{M}_{fi}=
\bar{u}({\bm p}_{Y})\,\sum_{i=1}^{6}\, 
A_{i}\, M_{i}\,\, u({\bm p}_{N})
 ~=~ \left(\frac{E_{N} + m_{N}}{2m_{N}} \right)^{{1}/{2}}
        \left(\frac{E_{Y} + m_{Y}}{2m_{Y}} 
        \right)^{{1}/{2}}}\nonumber\\ && \times\, \chi_{Y}^{\dagger}
        \, \Bigl(\, {\cal{F}}_{1}\,{{\bm \sigma} \,{\bm \cdot}\,} {\bm \epsilon}
+ {\cal{F}}_{2}\,  {\bm{\sigma}}{\bm \,\cdot\,} {\bm k}\, \epsilon_{0}
+ {\cal{F}}_{3}\, {\bm{\sigma}}\bm{\, \cdot \,}{\bm k}\, 
{\bm k}\bm{\, \cdot \,} {\bm{\epsilon}} + {\cal{F}}_{4}\, 
{\bm{\sigma}}\bm{\, \cdot \,}{\bm k}\, {\bm p}_{N}\bm{\, \cdot \,}
{\bm{\epsilon}}  + {\cal{F}}_{5}\, {\bm{\sigma}\,\cdot\,} {\bm k}\, {\bm p}_{Y}\bm{\, \cdot \,}{\bm{\epsilon}}\nonumber\\ 
&& 
+ {\cal{F}}_{6}\, {\bm{\sigma}\,\cdot\,} {\bm p}_{N}\,\epsilon_{0} 
+ {\cal{F}}_{7}\, {\bm{\sigma}} 
\cdot {\bm p}_{N}\, {\bm k}\bm{\, \cdot \,}{\bm{\epsilon}}
+ {\cal{F}}_{8}\, {\bm{\sigma}}\bm\, \cdot \,{\bm p}_{N}\, {\bm p}_{N} \,{\bm \cdot}\, 
{\bm {\epsilon}} + {\cal{F}}_{9}\, 
{\bm{\sigma}}\bm{\, \cdot \,}{\bm p}_{N}\, 
{\bm p}_{Y}\bm{\, \cdot \,} {\bm{\epsilon}} \nonumber\\
&& 
+ {\cal{F}}_{10}\, {\bm{\sigma}}\bm{\, \cdot \,}{\bm p}_{Y}\, 
\epsilon_{0} 
+ {\cal{F}}_{11}\, {\bm{\sigma}}\bm{\, \cdot \,}{\bm p}_{Y}\, {\bm k}\bm{\, \cdot \,}
{\bm{\epsilon}} 
+{\cal{F}}_{12}\, {\bm{\sigma}}
\bm{\, \cdot \,}{\bm p}_{Y}\, {\bm p}_{N}\bm{\, \cdot \,}{\bm{\epsilon}} + 
{\cal{F}}_{13}\, {\bm{\sigma}}\bm{\, \cdot \,}{\bm p}_{Y}\, 
{\bm p}_{Y}\bm{\, \cdot \,} {\bm{\epsilon}}
\nonumber\\ 
&& 
 + {\cal{F}}_{14}\, {\bm{\sigma}}\bm{\, \cdot \,}
{\bm{\epsilon}}\, {\bm{\sigma}}
\bm{\, \cdot \,}{\bm k}\, {\bm{\sigma}}\bm{\, \cdot \,}{\bm p}_{N}+ {\cal{F}}_{15}\, {\bm{\sigma}}\bm{\, \cdot \,}{\bm p}_{Y}\, 
{\bm{\sigma}}\bm{\, \cdot \,}{\bm{\epsilon}}\, {\bm{\sigma}}\bm{\, \cdot \,}{\bm k}
 + {\cal{F}}_{16}\, {\bm{\sigma}}\bm{\, \cdot \,}{\bm p}_{Y}\, 
{\bm{\sigma}}\bm{\, \cdot \,}{\bm{\epsilon}}\, {\bm{\sigma}}\bm{\, \cdot \,}{\bm p}_{N} 
\nonumber\\ 
&& 
+ {\cal{F}}_{17}\, {\bm{\sigma}}\bm{\, \cdot \,}{\bm p}_{Y}\, 
 {\bm{\sigma}}\bm{\, \cdot \,}
{\bm k}\, {\bm{\sigma}}\bm{\, \cdot \,}{\bm p}_{N}\, \epsilon_{0} 
+ {\cal{F}}_{18}\, {\bm{\sigma}}\bm{\, \cdot \,}{\bm p}_{Y}\, 
{\bm{\sigma}}\bm{\, \cdot \,}{\bm k}\, {\bm{\sigma}\,\cdot\, } 
{\bm p}_{N}\, {\bm k}\bm{\, \cdot \,}{\bm{\epsilon}}
\nonumber\\ 
 & &  + {\cal{F}}_{19}\, {\bm{\sigma}}\bm{\, \cdot \,}{\bm p}_{Y}\, 
{\bm{\sigma}}\bm{\, \cdot \,}{\bm k}\, {\bm{\sigma}\,
\cdot\,} {\bm p}_{N}\, {\bm p}_{N}\bm{\, \cdot \,} {\bm{\epsilon}}
 +\, {\cal{F}}_{20}\, {\bm{\sigma}}\bm{\, \cdot \,}{\bm p}_{Y}\, 
{\bm{\sigma}}\bm{\, \cdot \,}{\bm k}\, {\bm{\sigma}} 
\cdot {\bm p}_{N}\, {\bm p}_{Y}\bm{\, \cdot \,} {\bm{\epsilon}}
\, \Bigr) \, \chi_{N} ~,
\label{eq:gen_nonrel_op}
\end{eqnarray}
where we have defined the amplitudes ${\cal F}_i$ that can be related to the amplitudes $A_i$ in Eq.~(\ref{eq:Ai_Mi}) by using the lengthy equations given in Appendix A of Ref.~\cite{Mart:2008gq}.

In the literature, it is customary to recast the operator in Eq.~(\ref{eq:gen_nonrel_op}) into the spin–non-flip and spin–flip amplitudes, $L$ and $\bm{K}$, respectively, as
\begin{eqnarray}
    \label{eq:L_and_K}
\mathcal{M}_{fi} = \chi_{Y}^{\dagger}\, \bigl(L + i\,\bm{\sigma}\cdot\bm{K}\bigr)\,\chi_{N},
\end{eqnarray}
with \cite{Mart:1996ay}
\begin{eqnarray}
L & = & N~ \Bigl\{ -\left( {\cal{F}}_{14}+{\cal{F}}_{15}-{\cal{F}}_{16}
\right)~ {\bm p}_N \cdot ({\bm k} \times {\bm{\epsilon}}\, ) 
+{\cal{F}}_{15}~{\bm q}_K\cdot ({\bm k} \times {\bm{\epsilon}\, })
\nonumber\\ && \hspace{7mm} - {\cal{F}}_{16}~{\bm p}_N \cdot 
({\bm q}_K \times {\bm{\epsilon}\, })
-\left[ {\cal{F}}_{17}~\epsilon_0 + \left({\cal{F}}_{18}+{\cal{F}}_{20}
\right)~ {\bm k} \cdot {\bm{\epsilon}} \right.
\nonumber\\ 
&& \hspace{7mm} \left.
 +\left( {\cal{F}}_{19}+{\cal{F}}_{20} \right)~ {\bm p}_N \cdot 
{\bm{\epsilon}} - {\cal{F}}_{20}~ {\bm q}_K \cdot {\bm k}
\right]~ {\bm p}_N \cdot ({\bm q}_K \times {\bm{\epsilon}\, })
\Bigr\} ~,\\
  {\bm K} &=& -N \left( T_1~{\bm{\epsilon}}+T_2~{\bm k}+
T_3~{\bm p}_N+  T_4~{\bm q}_K \right) ~,
\end{eqnarray}
where
\begin{eqnarray}
\label{eq:normalization}
N &=& \left(\frac{E_{N} + m_{N}}{2m_{N}} \right)^{\frac{1}{2}}
\left(\frac{E_{Y} + m_{Y}}{2m_{Y}} \right)^{\frac{1}{2}} ~ ,
\end{eqnarray}
and
\begin{eqnarray}
  T_1 &=& {\cal{F}}_{1}+({\cal{F}}_{14}-{\cal{F}}_{15}-{\cal{F}}_{16})~
          {\bm p}_N \cdot {\bm k}+ {\cal{F}}_{15}~ ({\bm q}_K\cdot{\bm k}
          - {\bm k}^2) \nonumber\\ &&
          +{\cal{F}}_{16}~({\bm p}_N\cdot{\bm q}_K - {\bm p}_N^{\, 2}), \\[0.5ex]
  T_2 &=& [{\cal{F}}_{2}+{\cal{F}}_{10}+({\bm p}_N\cdot{\bm q}- 
     {\bm p}_N^{\, 2})~{\cal{F}}_{17}]~\epsilon_0+[{\cal{F}}_{3}+{\cal{F}}_{5}
      +{\cal{F}}_{11}+{\cal{F}}_{13}+2{\cal{F}}_{15}\nonumber\\
      && +({\bm p}_N \cdot {\bm q}-{\bm p}_N^{\, 2})~({\cal{F}}_{18}+
      {\cal{F}}_{20})]~{\bm k}\cdot{\bm{\epsilon}} 
      +[{\cal{F}}_{4}+
      {\cal{F}}_{5}+{\cal{F}}_{12}+{\cal{F}}_{13}-{\cal{F}}_{14}
      \nonumber\\
      && +{\cal{F}}_{15}+{\cal{F}}_{16}+({\bm p}_N \cdot {\bm q}-
      {\bm p}_N^{\, 2})~({\cal{F}}_{19}+
      {\cal{F}}_{20})]~{\bm p}_N\cdot{\bm{\epsilon}}
      -[{\cal{F}}_{5}+{\cal{F}}_{13}+{\cal{F}}_{15}\nonumber\\
      && +({\bm p}_N\cdot{\bm q}-{\bm p}_N^{\, 2})~
      {\cal{F}}_{20}]~{\bm q}_K\cdot{\bm{\epsilon}} ~,\\[0.5ex]
  T_3 &=& [{\cal{F}}_{6}+{\cal{F}}_{10}+(2{\bm p}_N\cdot{\bm k}+ 
      {\bm k}^2-{\bm q}_K\cdot{\bm k})~{\cal{F}}_{17}]~\epsilon_0+
      [{\cal{F}}_{7}+{\cal{F}}_{9}+{\cal{F}}_{11}+{\cal{F}}_{13}\nonumber\\
      &&+{\cal{F}}_{14}+{\cal{F}}_{15}+{\cal{F}}_{16} +(2{\bm p}_N \cdot 
      {\bm k}+{\bm k}^2-{\bm q}_K\cdot{\bm k})~ ({\cal{F}}_{18}+
      {\cal{F}}_{20})]~{\bm k}\cdot{\bm{\epsilon}}
      + [{\cal{F}}_{8} \nonumber\\
      &&+{\cal{F}}_{9}+{\cal{F}}_{12}+{\cal{F}}_{13}+2{\cal{F}}_{16}
      +(2{\bm p}_N \cdot {\bm k}+{\bm k}^2-{\bm q}_K\cdot{\bm k})~
     ({\cal{F}}_{19}+{\cal{F}}_{20})]~{\bm p}_N\cdot
      {\bm{\epsilon}}\nonumber\\
      && -[{\cal{F}}_{9}+{\cal{F}}_{13}+{\cal{F}}_{16}
      +(2{\bm p}_N \cdot {\bm k}+{\bm k}^2-{\bm q}_K\cdot{\bm k})~
      {\cal{F}}_{20}]~{\bm q}_K\cdot{\bm{\epsilon}} ~,\\[0.5ex]
  T_4 &=& -({\cal{F}}_{10}+{\cal{F}}_{17}~{\bm p}_N \cdot {\bm k})~
      \epsilon_0-[{\cal{F}}_{11}+{\cal{F}}_{13}+{\cal{F}}_{15}+
      ({\cal{F}}_{18}+{\cal{F}}_{20})~{\bm p}_N \cdot {\bm k}]~
      {\bm k}\cdot{\bm{\epsilon}}\nonumber\\
      &&+[{\cal{F}}_{12}+{\cal{F}}_{13}+{\cal{F}}_{16}+
      ({\cal{F}}_{19}+{\cal{F}}_{20})~{\bm p}_N \cdot {\bm k}]~
      {\bm p}_N\cdot{\bm{\epsilon}} \nonumber\\
      &&+({\cal{F}}_{13}+{\cal{F}}_{20}~
      {\bm p}_N\cdot{\bm k})~{\bm q}_K\cdot{\bm{\epsilon}} ~.
\end{eqnarray}

In this representation the spin–independent ($L$) and spin–dependent ($\bm{K}$) components of the operator are explicitly separated, making it particularly convenient for nuclear applications. The amplitude expressed in Eq.~(\ref{eq:L_and_K}) was employed, for instance, in Ref.~\cite{Mart:1996ay} to investigate the photoproduction of hypertriton, and in Ref.~\cite{Yamamura:1999xm} to investigate the $\Lambda$- and $\Sigma$-threshold phenomena in both inclusive $K^+$ and exclusive $K^+Y$ photoproduction on the deuteron.

At this stage, it is important to emphasize that both the spin–non-flip ($L$) and spin–flip ($\bm{K}$) amplitudes depend on the photon polarization vector $\bm{\epsilon}$. This dependence may complicate calculations in nuclear applications, since the polarization vector is typically defined in the nuclear rest frame. To address this issue, the amplitude in Eq.~(\ref{eq:L_and_K}) can be rewritten as~\cite{Miyagawa:2006kj}
\begin{eqnarray}
    \mathcal{M}_{fi} = \chi_{Y}^{\dagger}~ t_{\gamma K} ~ \chi_{N}\, ,
\end{eqnarray}
with
\begin{eqnarray} 
\label{eq:t_gamma_K} 
t_{\gamma K} &=& (1,i\sigma_x,i\sigma_y,i\sigma_z) \,
\left( \begin{array}{ccc}
{L}_x & {L}_y & {L}_z \\ {K}_{xx} & {K}_{xy} & {K}_{xz} \\ 
{K}_{yx} & {K}_{yy} & {K}_{yz} \\ 
{K}_{zx} & {K}_{zy} & {K}_{zz} \end{array}\right) ~
\left( \begin{array}{c}
\epsilon_x \\ \epsilon_y \\ \epsilon_z  \end{array}\right) ~.
\end{eqnarray}

Equation~(\ref{eq:t_gamma_K}) shows that the elementary operator is now expressed in the form of a $4\times 3$ matrix, which is completely independent of the reference frame in which the spin operator $\boldsymbol{\sigma}$ and the photon polarization vector $\boldsymbol{\epsilon}$ are defined. In contrast to Eq.~(\ref{eq:L_and_K}), the spin–non-flip amplitude is now represented as a vector~\cite{Miyagawa:2006kj},
\begin{eqnarray} 
\bm{L} &=& N \Bigl[ {\cal F}_{14}\,\bm{k} \times \bm{p}_N +
{\cal F}_{15}\,\bm{k} \times \bm{p}_Y +{\cal F}_{16}\,\bm{p}_N
 \times \bm{p}_Y \nonumber\\ &&
 +~ \bm{p}_Y\bm{\,\cdot\, k} \times \bm{p}_N \bigl({\cal F}_{18}\, \bm{k} +
{\cal F}_{19}\, \bm{p}_N + {\cal F}_{20}\, \bm{p}_Y\bigr)\Bigr] ,
\end{eqnarray}
whereas the spin–flip amplitude is expressed as a tensor,
\begin{eqnarray} 
{K}_{ij} &=& \delta_{ij}\,A + p_{\gamma ,i}\,{B}_j + 
p_{N,i}\,{C}_j + p_{Y,i}\,{D}_j  ~, ~~~~ i,j=x,y,z  ~~,
\end{eqnarray}
with 
\begin{eqnarray} 
A &=& -N \left[{\cal F}_1+{\cal F}_{14}\,\bm{p}_N\bm{\,\cdot\, k} -
{\cal F}_{15}\,\bm{p}_Y\bm{\,\cdot\,k}-{\cal F}_{16}\,\bm{p}_N
\bm{\,\cdot\,p}_Y\right]
~,\\
\bm{B} &=& -N \left[({\cal F}_4-{\cal F}_{14}-{\cal F}_{19}\,
\bm{p}_N\bm{\,\cdot\, p}_Y)\,\bm{p}_N 
+ ({\cal F}_{5}+{\cal F}_{15}-{\cal F}_{20}\,\bm{p}_N\bm{\,\cdot\, p}_Y)\,
\bm{p}_Y 
\right] \, ,~~\\
\bm{C} &=& -N \left[({\cal F}_8+{\cal F}_{19}\,\bm{p}_Y\bm{\,\cdot\, p}
_\gamma)\,\bm{p}_N 
+ ({\cal F}_{9}+{\cal F}_{16}+{\cal F}_{20}\,\bm{p}_Y
\bm{\,\cdot\, k})\,\bm{p}_Y \right] ~,
\\
\bm{D} &=& -N \left[({\cal F}_{12}+{\cal F}_{16}+{\cal F}_{19}\,\bm{p}_N
\bm{\,\cdot\, k})\,\bm{p}_N 
+ ({\cal F}_{13}+{\cal F}_{20}\,\bm{p}_N\bm{\,\cdot\,k})\,\bm{p}_Y
 \right]
~,
\end{eqnarray}
whereas $N$ is given by 
\begin{equation} 
N= \left(\frac{E_N+m_N}{2m_N}\right)^{{1}/{2}}
\left(\frac{E_Y+m_Y}{2m_Y}\right)^{{1}/{2}}
\sqrt{\frac{m_Y}{E_Y}}
\sqrt{\frac{m_N}{E_N}} ~.
\end{equation}
The amplitude given in Eq.~(\ref{eq:t_gamma_K}) was employed, for example, in Refs.~\cite{Miyagawa:2006kj,Salam:2006kk} to investigate kaon photoproduction on the deuteron, both in the quasi-free scattering region, where one nucleon acts as a spectator with negligible momentum, and in the kinematical region where the hyperon–nucleon FSI plays a significant role.

Finally, the elementary operator can also be written as
\begin{eqnarray}
\mathcal{M}_{\mathrm{fi}} = \langle\, {\rm f}\, |\, \epsilon_\mu\, J^\mu\, |\, {\rm i} \,\rangle .
\end{eqnarray}
Since $\epsilon_\mu$ is defined in the nuclear rest frame, it is necessary to evaluate the term in the elementary operator that transforms the nucleon state with spin projection $m_s$ into the hyperon state with spin projection $m_{s'}$, i.e., $\langle \tfrac{1}{2}\, m_{s'}\, |\, J^\mu\, |\, \tfrac{1}{2}\, m_s \rangle$. For this purpose, $J^\mu$ can be rewritten as \cite{Mart:2008gq} 
\begin{eqnarray}
J^\mu &=& (1,\sigma_x,\sigma_y,\sigma_z)
\left(\begin{array}{cccc}
j_{00} & j_{x0} & j_{y0} & j_{z0}\\
j_{0x} & j_{xx} & j_{yx} & j_{zx}\\
j_{0y} & j_{xy} & j_{yy} & j_{zy}\\
j_{0z} & j_{xz} & j_{yz} & j_{zz}
\end{array} \right) \nonumber\\
&=& j^\mu_0 + \sigma_x\, j^\mu_x + \sigma_y\, j^\mu_y + \sigma_z\, j^\mu_z \nonumber\\
&=& \sum_{n=0}^1~\sum_{m_n=-n}^{+n}\,
(-1)^{m_n}\,\sigma^{(n)}_{-m_n}\,
[\,j^{\mu}\,]^{(n)}_{m_n} ,
\label{eq:define_j}
\end{eqnarray}
which is particularly convenient because the spin operator $\sigma^{(n)}_{-m_n}$ can be evaluated separately, i.e.,
\begin{eqnarray}
\langle {\textstyle \frac{1}{2}} , m_{s'}\, |\,\sigma_{-m_n}^{(n)}\, |\,
{\textstyle \frac{1}{2}},m_{s}\rangle &=&
\sqrt{2}\, (-1)^{n-\frac{1}{2}-m_{s'}+m_{n}}
\left( \tfrac{1}{2}-\! m_{s'}\,\tfrac{1}{2} m_{s}\,\middle|\, n m_{n} \right) ,
\label{eq:bracket1}
\end{eqnarray}
and the resulting operator $[\,j^{\mu}\,]^{(n)}_{m_n}$ is completely frame independent. Note that in Eq.~(\ref{eq:bracket1}) we have defined
\begin{equation*}
     \sigma^{(0)} = 1, ~~~~~ \sigma^{(1)} = \boldsymbol{\sigma} , 
\end{equation*}
and
\begin{equation*}
     \left[\, j^{\mu}\, \right]^{(1)}_{\pm 1} = \mp\frac{1}{\sqrt{2}}(j^{\mu}_x\pm ij^{\mu}_y), ~~~~~~ \left[\, j^{\mu}\, \right]^{(1)}_{0} = j^{\mu}_z, ~~~~~~  \left[\, j^{\mu}\, \right]^{(0)} = j^\mu_0 .
\end{equation*}

The components of the matrix defined in Eq.~(\ref{eq:define_j}) are given by \cite{Mart:2008gq}
\begin{eqnarray}
  j_{00} &=& iN{\cal{F}}_{17}\,\bm{p}_{N} 
    \cdot (\bm{p}_Y \times \bm{k})~,\\
  j_{\ell 0} &=& iN \Bigl[{\cal F}_{14}\, \bm{p}_N \times \bm{k}+
{\cal F}_{15}\,\bm{p}_Y \times \bm{k} -{\cal F}_{16}\,\bm{p}_N
 \times \bm{p}_Y \nonumber\\ && 
~~~~~~- \bm{p}_N\cdot\bm{p}_Y \times \bm{k} 
\left({\cal F}_{18}\, \bm{k} + {\cal F}_{19}\, \bm{p}_N + {\cal F}_{20}\, \bm{p}_Y\right)\Bigr]_\ell~, \\
  j_{0m} &=& N \Bigl\{({\cal F}_2- \bm{p}_{N} 
    \cdot \bm{p}_Y\,{\cal F}_{17})\,k_m+
        ({\cal F}_6+ \bm{p}_{Y} 
    \cdot \bm{k}\,{\cal F}_{17})\,p_{N,m}\nonumber\\
    && ~~~~~~+ ({\cal F}_{10}+ \bm{p}_{N} 
    \cdot \bm{k}\,{\cal F}_{17})\,p_{Y,m}
        \Bigr\} ,~~~~\\
  j_{\ell m} &=& A\,\delta_{\ell m} + {B}_{\ell}\,k_m + 
  {C}_\ell\,p_{N,m}  + {D}_\ell\,p_{Y,m}  ~,
\end{eqnarray}
where $\ell,m=x,y,z$, and 
\begin{eqnarray} 
A &=& -N \left[{\cal F}_1+{\cal F}_{14}\,\bm{p}_N\cdot\bm{k} -
{\cal F}_{15}\,\bm{p}_Y\cdot\bm{k}-{\cal F}_{16}\,\bm{p}_N
\cdot\bm{p}_Y\right]
~,\\[0.5ex]
\bm{B} &=& -N \left[({\cal{F}}_{3}-
  \bm{p}_{N}\cdot\bm{p}_Y\; {\cal{F}}_{18})\; \bm{k}
  + ( {\cal{F}}_{4}-{\cal{F}}_{14}-\bm{p}_{N}\cdot\bm{p}_Y\;
  {\cal{F}}_{19})\;\bm{p}_{N}\right. \nonumber\\
  && \hspace{22mm}\left.
  +~ ( {\cal{F}}_{5}+{\cal{F}}_{15}-\bm{p}_{N}\cdot\bm{p}_Y\;
  {\cal{F}}_{20})\;\bm{p}_{Y}
\right] ~,\\[0.5ex]
\bm{C} &=& -N \left[ ({\cal{F}}_{7}+{\cal F}_{14} + 
  \bm{p}_{Y}\cdot\bm{k}\; {\cal{F}}_{18})\; \bm{k}
  + ( {\cal{F}}_{8}+\bm{p}_{Y}\cdot\bm{k}\;
  {\cal{F}}_{19})\;\bm{p}_{N}\right. \nonumber\\
  && \hspace{22mm}\left.
  +~ ( {\cal{F}}_{9}+{\cal{F}}_{16}+\bm{p}_{Y}\cdot\bm{k}\;
  {\cal{F}}_{20})\;\bm{p}_{Y}
 \right] ~,
\\[0.5ex]
\bm{D} &=& -N \left[ ({\cal{F}}_{11}+{\cal F}_{15} + 
  \bm{p}_{N}\cdot\bm{k}\; {\cal{F}}_{18})\; \bm{k}
  +~ ( {\cal{F}}_{12}+{\cal F}_{16}+\bm{p}_{N}\cdot\bm{k}\;
  {\cal{F}}_{19})\;\bm{p}_{N}\right. \nonumber\\
  && \hspace{22mm}\left.
  +~ ( {\cal{F}}_{13}+\bm{p}_{N}\cdot\bm{k}\;
  {\cal{F}}_{20})\;\bm{p}_{Y}
 \right]
~,
\end{eqnarray}
with $N$ is given by Eq.~(\ref{eq:normalization}).

The amplitude given in Eq.~(\ref{eq:define_j}) was formulated in Ref.~\cite{Mart:2008gq} to investigate kaon photoproduction on $^3$He leading to the formation of the hypertriton $^3_\Lambda$H in the final state. This formalism is particularly suitable for application within the partial-wave representation, in which the wave functions of both the initial and final bound states are expanded in terms of partial waves.

\subsection{Verifying the Operator’s Numerical Output}
After incorporating the elementary operator into the nuclear environment, it is important to verify the numerical output for consistency. This verification is performed within the nuclear calculation framework by reducing the nuclear system to a single-nucleon system, so that the numerical result reproduces the elementary process, i.e., kaon photoproduction on the nucleon. For example, in Ref.~\cite{Mart:2008gq}, the nuclear amplitude was reduced to the elementary amplitude by replacing the nuclear wave functions with unity and retaining only the necessary summations over angular momenta and spin projections. As a result, the nuclear scattering amplitude, given by Eq.~(33) in Ref.~\cite{Mart:2008gq}, reduces to
\begin{eqnarray}
  \langle\, {\rm f} \left| \, J^\mu \, \right| {\rm i} \,\rangle &=& 
  \sqrt{2}\, \sum_{n,m_{n}} (-1)^{n-1/2-M_{\rm f}}
        \left({\textstyle \frac{1}{2}} \;-\!\!M_{\rm f}\; 
        {\textstyle \frac{1}{2}}\; M_{\rm i}\,|\, n\, m_n\right)
   \,\left[\,j^{\mu}\,\right]^{(n)}_{m_n}
    \, ,
\label{trans_elementary}
\end{eqnarray}
from which the transverse and longitudinal cross sections can be written as ~\cite{Mart:2008gq}
\begin{eqnarray}
\label{cs_elementary1}
  \frac{d\sigma_{\rm T}}{d\Omega_K^{\rm c.m.}} &=& 
  \frac{q_{K}^{\rm c.m.}}{k^{\rm c.m.}}\, \frac{m_pm_\Lambda}{32\pi^2W^2}\, 
        \sum_{i=0}^3 \left( |j_{ix}|^2+|j_{iy}|^2\right) ~ ,\\
  \frac{d\sigma_{\rm L}}{d\Omega_K^{\rm c.m.}} &=& 
  \frac{q_{K}^{\rm c.m.}}{k^{\rm c.m.}}\, \frac{m_pm_\Lambda}{32\pi^2W^2}\, 
        \sum_{i=0}^3  2|j_{i0}|^2 ~ ,
\label{cs_elementary2}
\end{eqnarray}
and are found to be identical to the standard definitions of the transverse and longitudinal cross sections in the elementary process. It should be emphasized that Eqs.~(\ref{cs_elementary1}) and (\ref{cs_elementary2}) are provided only for analytical comparison. The actual evaluation must be carried out numerically using the nuclear calculation code, with the appropriate reductions described above.

\section{Conclusion}
\label{sec:conclusion}
An elementary operator has been successfully constructed for application in kaon photoproduction on both the nucleon and nuclei. In the case of kaon photoproduction on the nucleon, the operator successfully reproduces the presently available experimental data in six isospin channels, including differential cross sections as well as single- and double-polarization observables. The fitting procedure was carried out using nearly 17,000 data points. For kaon photoproduction on nuclei, the operator has been formulated in a fully frame-independent manner. Furthermore, a general extension has been developed to accommodate the electroproduction process, which will be addressed in future work. Since kaon photoproduction on nuclei can be described within different theoretical frameworks, such as the impulse approximation, covariant formulations, or few-body approaches, the output of the elementary operator must likewise be expressed in corresponding representations suitable for each framework. To this end, five alternative formulations for expressing the amplitude have been proposed.

\begin{acknowledgements}
This work has been supported by the PUTI Q1 Research Grant from the University of Indonesia under contract No. NKB 442/UN2.RST/HKP.05.00/2024.
\end{acknowledgements}

\bibliographystyle{spmpsci}
\bibliography{ref}   

\begin{thebibliography}{10}
\providecommand{\url}[1]{{#1}}
\providecommand{\urlprefix}{URL }
\expandafter\ifx\csname urlstyle\endcsname\relax
  \providecommand{\doi}[1]{DOI~\discretionary{}{}{}#1}\else
  \providecommand{\doi}{DOI~\discretionary{}{}{}\begingroup \urlstyle{rm}\Url}\fi

\bibitem{Moravcsik:1957}
Kawaguchi, M., Moravcsik, M.J.: {Photoproduction of $K$ mesons from single nucleons}.
\newblock Phys. Rev. \textbf{107}, 563 (1957).
\newblock \doi{10.1103/PhysRev.107.563}

\bibitem{Thom:1966rm}
Thom, H.: {Phenomenological analysis of $K^+\Lambda$ photoproduction}.
\newblock Phys. Rev. \textbf{151}, 1322--1336 (1966).
\newblock \doi{10.1103/PhysRev.151.1322}

\bibitem{Adelseck:1985scp}
Adelseck, R.A., Bennhold, C., Wright, L.E.: {Kaon photoproduction operator for use in nuclear physics}.
\newblock Phys. Rev. C \textbf{32}, 1681--1692 (1985).
\newblock \doi{10.1103/PhysRevC.32.1681}

\bibitem{Adelseck:1990ch}
Adelseck, R.A., Saghai, B.: {Kaon photoproduction: Data consistency, coupling constants, and polarization observables}.
\newblock Phys. Rev. C \textbf{42}, 108--127 (1990).
\newblock \doi{10.1103/PhysRevC.42.108}

\bibitem{Mart:1995wu}
Mart, T., Bennhold, C., Hyde-Wright, C.E.: {Constraints on coupling constants through charged $\Sigma$ photoproduction}.
\newblock Phys. Rev. C \textbf{51}, R1074--R1077 (1995).
\newblock \doi{10.1103/PhysRevC.51.R1074}

\bibitem{David:1995pi}
David, J.C., Fayard, C., Lamot, G.H., Saghai, B.: {Electromagnetic production of associated strangeness}.
\newblock Phys. Rev. C \textbf{53}, 2613--2637 (1996).
\newblock \doi{10.1103/PhysRevC.53.2613}

\bibitem{Mart:1999ed}
Mart, T., Bennhold, C.: {Evidence for a missing nucleon resonance in kaon photoproduction}.
\newblock Phys. Rev. C \textbf{61}, 012,201 (2000).
\newblock \doi{10.1103/PhysRevC.61.012201}

\bibitem{Maxwell:2004ga}
Maxwell, O.V.: {Model dependence in the photoproduction of kaons from protons and deuterons}.
\newblock Phys. Rev. C \textbf{70}, 044,612 (2004).
\newblock \doi{10.1103/PhysRevC.70.044612}

\bibitem{Mart:2019mtq}
Mart, T.: {Coupled $K^+\Lambda$ and $K^0\Lambda$ photoproduction off the nucleon: Consequences from the recent CLAS and MAMI data and the $N(1680)P_{11}$ narrow state}.
\newblock Phys. Rev. D \textbf{100}(5), 056,008 (2019).
\newblock \doi{10.1103/PhysRevD.100.056008}

\bibitem{SAPHIR:1998fev}
Tran, M.Q., et~al.: {Measurement of $\gamma p \to K^+\Lambda$ and $\gamma p \to K^0\Sigma^+$ at photon energies up to 2 GeV}.
\newblock Phys. Lett. B \textbf{445}, 20--26 (1998).
\newblock \doi{10.1016/S0370-2693(98)01393-8}

\bibitem{SAPHIR:1999wfu}
Goers, S., et~al.: {Measurement of $\gamma p \to K^0\Sigma^+$ at photon energies up to 1.55 GeV}.
\newblock Phys. Lett. B \textbf{464}, 331--338 (1999).
\newblock \doi{10.1016/S0370-2693(99)01031-X}

\bibitem{CLAS:2005lui}
Bradford, R., et~al.: {Differential cross sections for $\gamma + p \to K^+ + Y$ for $\Lambda$ and $\Sigma^0$ hyperons}.
\newblock Phys. Rev. C \textbf{73}, 035,202 (2006).
\newblock \doi{10.1103/PhysRevC.73.035202}

\bibitem{CLAS:2006pde}
Bradford, R.K., et~al.: {First measurement of beam-recoil observables $C_x$ and $C_z$ in hyperon photoproduction}.
\newblock Phys. Rev. C \textbf{75}, 035,205 (2007).
\newblock \doi{10.1103/PhysRevC.75.035205}

\bibitem{CBELSATAPS:2007oqn}
Castelijns, R., et~al.: {Nucleon resonance decay by the $K^0 \Sigma^+$ channel}.
\newblock Eur. Phys. J. A \textbf{35}, 39--45 (2008).
\newblock \doi{10.1140/epja/i2007-10529-8}

\bibitem{GRAAL:2008jrm}
Lleres, A., et~al.: {Measurement of beam-recoil observables $O_x,~O_z$ and target asymmetry for the reaction $\gamma p \to K^+ \Lambda$}.
\newblock Eur. Phys. J. A \textbf{39}, 149--161 (2009).
\newblock \doi{10.1140/epja/i2008-10713-4}

\bibitem{CLAS:2009rdi}
McCracken, M.E., et~al.: {Differential cross section and recoil polarization measurements for the $\gamma p \to K^+ \Lambda$ reaction using CLAS at Jefferson Lab}.
\newblock Phys. Rev. C \textbf{81}, 025,201 (2010).
\newblock \doi{10.1103/PhysRevC.81.025201}

\bibitem{CLAS:2009fmu}
Pereira, S.A., et~al.: {Differential cross section of $\gamma n \to K^+ \Sigma^-$ on bound neutrons with incident photons from 1.1 to 3.6 GeV}.
\newblock Phys. Lett. B \textbf{688}, 289--293 (2010).
\newblock \doi{10.1016/j.physletb.2010.04.028}

\bibitem{CLAS:2010aen}
Dey, B., et~al.: {Differential cross sections and recoil polarizations for the reaction $\gamma p \to K^{+} \Sigma^{0}$}.
\newblock Phys. Rev. C \textbf{82}, 025,202 (2010).
\newblock \doi{10.1103/PhysRevC.82.025202}

\bibitem{CrystalBallatMAMI:2013iig}
Jude, T.C., et~al.: {$K^+\Lambda$ and $K^+\Sigma^0$ photoproduction with fine center-of-mass energy resolution}.
\newblock Phys. Lett. B \textbf{735}, 112--118 (2014).
\newblock \doi{10.1016/j.physletb.2014.06.015}

\bibitem{CLAS:2016wrl}
Paterson, C.A., et~al.: {Photoproduction of $\Lambda$ and $\Sigma^0$ hyperons using linearly polarized photons}.
\newblock Phys. Rev. C \textbf{93}(6), 065,201 (2016).
\newblock \doi{10.1103/PhysRevC.93.065201}

\bibitem{CLAS:2017gsu}
Compton, N., et~al.: {Measurement of the differential and total cross sections of the ${\gamma}d{\rightarrow}{K}^{0}\mathrm{{\Lambda}}(p)$ reaction within the resonance region}.
\newblock Phys. Rev. C \textbf{96}(6), 065,201 (2017).
\newblock \doi{10.1103/PhysRevC.96.065201}

\bibitem{Mart:2010ch}
Mart, T.: {Electromagnetic production of kaon near threshold}.
\newblock Phys. Rev. C \textbf{82}, 025,209 (2010).
\newblock \doi{10.1103/PhysRevC.82.025209}

\bibitem{Haberzettl:1998aqi}
Haberzettl, H., Bennhold, C., Mart, T., Feuster, T.: {Gauge-invariant tree-level photoproduction amplitudes with form factors}.
\newblock Phys. Rev. C \textbf{58}(1), R40--R44 (1998).
\newblock \doi{10.1103/PhysRevC.58.R40}

\bibitem{Pascalutsa:2000kd}
Pascalutsa, V.: {Correspondence of consistent and inconsistent spin - 3/2 couplings via the equivalence theorem}.
\newblock Phys. Lett. B \textbf{503}, 85--90 (2001).
\newblock \doi{10.1016/S0370-2693(01)00140-X}

\bibitem{Vrancx:2011qv}
Vrancx, T., De~Cruz, L., Ryckebusch, J., Vancraeyveld, P.: {Consistent interactions for high-spin fermion fields}.
\newblock Phys. Rev. C \textbf{84}, 045,201 (2011).
\newblock \doi{10.1103/PhysRevC.84.045201}

\bibitem{Luthfiyah:2021yqe}
Luthfiyah, N.H., Mart, T.: {Role of the high-spin nucleon and delta resonances in the $K\Lambda$ and $K\Sigma$ photoproduction off the nucleon}.
\newblock Phys. Rev. D \textbf{104}, 076,022 (2021).
\newblock \doi{10.1103/PhysRevD.104.076022}

\bibitem{Mart:2015jof}
Mart, T., Clymton, S., Arifi, A.J.: {Nucleon resonances with spin 3/2 and 5/2 in the isobar model for kaon photoproduction}.
\newblock Phys. Rev. D \textbf{92}(9), 094,019 (2015).
\newblock \doi{10.1103/PhysRevD.92.094019}

\bibitem{Mart:2008gq}
Mart, T., Van Der~Ventel, B.: {Photo- and electroproduction of the hypertriton on He-3}.
\newblock Phys. Rev. C \textbf{78}, 014,004 (2008).
\newblock \doi{10.1103/PhysRevC.78.014004}

\bibitem{Knochlein:1995qz}
Knochlein, G., Drechsel, D., Tiator, L.: {Photoproduction and electroproduction of eta mesons}.
\newblock Z. Phys. A \textbf{352}, 327--343 (1995).
\newblock \doi{10.1007/BF01289506}

\bibitem{Dennery:1961zz}
Dennery, P.: {Theory of the electro- and photoproduction of $\pi$ mesons}.
\newblock Phys. Rev. \textbf{124}, 2000--2010 (1961).
\newblock \doi{10.1103/PhysRev.124.2000}

\bibitem{Chew:1957tf}
Chew, G.F., Goldberger, M.L., Low, F.E., Nambu, Y.: {Relativistic dispersion relation Approach to photomeson production}.
\newblock Phys. Rev. \textbf{106}, 1345--1355 (1957).
\newblock \doi{10.1103/PhysRev.106.1345}

\bibitem{Clymton:2021wof}
Clymton, S., Mart, T.: {Extracting the pole and Breit-Wigner properties of nucleon and {\ensuremath{\Delta}} resonances from the $\gamma N\to K\Sigma$ photoproduction}.
\newblock Phys. Rev. D \textbf{104}(5), 056,015 (2021).
\newblock \doi{10.1103/PhysRevD.104.056015}

\bibitem{Bennhold:1999mt}
Bennhold, C., Haberzettl, H., Mart, T.: {A new resonance in $K^+ \Lambda$ electroproduction: The $D_{13}(1895)$ and its electromagnetic form-factors}.
\newblock In: {2nd ICTP International Conference on Perspectives in Hadronic Physics}, pp. 328--337 (1999)

\bibitem{LEPS:2005hji}
Sumihama, M., et~al.: {The polarized $\gamma p \to K^+ \Lambda$ and polarized $\gamma p \to K^+ \Sigma^0$ reactions at forward angles with photon energies from 1.5-GeV to 2.4-GeV}.
\newblock Phys. Rev. C \textbf{73}, 035,214 (2006).
\newblock \doi{10.1103/PhysRevC.73.035214}

\bibitem{A2:2018doh}
Akondi, C.S., et~al.: {Experimental study of the $\gamma p\rightarrow K^0\Sigma^+$, $\gamma n\rightarrow K^0\Lambda$, and $\gamma n\rightarrow K^0 \Sigma^0$ reactions at the Mainz Microtron}.
\newblock Eur. Phys. J. A \textbf{55}(11), 202 (2019).
\newblock \doi{10.1140/epja/i2019-12924-x}

\bibitem{Mart:2000jv}
Mart, T.: {Role of $P_{13}(1720)$ in $K\Sigma$ photoproduction}.
\newblock Phys. Rev. C \textbf{62}, 038,201 (2000).
\newblock \doi{10.1103/PhysRevC.62.038201}

\bibitem{Lawall:2005np}
Lawall, R., et~al.: {Measurement of the reaction $\gamma p \to K^0\Sigma^+$ at photon energies up to 2.6-GeV}.
\newblock Eur. Phys. J. A \textbf{24}, 275--286 (2005).
\newblock \doi{10.1140/epja/i2005-10002-x}

\bibitem{A2:2013cqk}
Aguar-Bartolome, P., et~al.: {Measurement of the $\gamma p \to K^{0} \Sigma^{+}$ reaction with the Crystal Ball/TAPS detectors at the Mainz Microtron}.
\newblock Phys. Rev. C \textbf{88}(4), 044,601 (2013).
\newblock \doi{10.1103/PhysRevC.88.044601}

\bibitem{Kohri:2006yx}
Kohri, H., et~al.: {Differential cross section and photon beam asymmetry for the polarized $\gamma n \to K^+ \Sigma^-$ reaction at $E_\gamma = 1.5$ GeV - 2.4 GeV}.
\newblock Phys. Rev. Lett. \textbf{97}, 082,003 (2006).
\newblock \doi{10.1103/PhysRevLett.97.082003}

\bibitem{Salam:2004gz}
Salam, A., Arenhovel, H.: {Interaction effects in $K^+$ photoproduction on the deuteron}.
\newblock Phys. Rev. C \textbf{70}, 044,008 (2004).
\newblock \doi{10.1103/PhysRevC.70.044008}

\bibitem{Mart:1996ay}
Mart, T., Tiator, L., Drechsel, D., Bennhold, C.: {Electromagnetic production of the hypertriton}.
\newblock Nucl. Phys. A \textbf{640}, 235--258 (1998).
\newblock \doi{10.1016/S0375-9474(98)00441-2}

\bibitem{Yamamura:1999xm}
Yamamura, H., Miyagawa, K., Mart, T., Bennhold, C., Gloeckle, W.: {Inclusive $K^+$ and exclusive $K^+Y$ photoproduction on the deuteron: $\Lambda$ and $\Sigma$ threshold phenomena}.
\newblock Phys. Rev. C \textbf{61}, 014,001 (2000).
\newblock \doi{10.1103/PhysRevC.61.014001}

\bibitem{Miyagawa:2006kj}
Miyagawa, K., Mart, T., Bennhold, C., Glockle, W.: {Polarization observables in exclusive kaon photoproduction on the deuteron}.
\newblock Phys. Rev. C \textbf{74}, 034,002 (2006).
\newblock \doi{10.1103/PhysRevC.74.034002}

\bibitem{Salam:2006kk}
Salam, A., Miyagawa, K., Mart, T., Bennhold, C., Glockle, W.: {$K^0$ photoproduction on the deuteron and the extraction of the elementary amplitude}.
\newblock Phys. Rev. C \textbf{74}, 044,004 (2006).
\newblock \doi{10.1103/PhysRevC.74.044004}

\end{thebibliography}

%
%

\end{document}